\documentclass[tightenlines,twoside,secnumarabic,onecolumn,floatfix,nofootinbib,preprintnumbers,11pt]{revtex4}

\usepackage{graphicx}
\usepackage{dcolumn}
\usepackage{float}
\usepackage{amsmath}
\usepackage{amsfonts}
\usepackage{multirow}
\usepackage{color}
\usepackage{hyperref}
\def\GeV{\ifmmode {\mathrm{\ Ge\kern -0.1em V}}\else
                   \textrm{Ge\kern -0.1em V}\fi}%
\newcommand\rG{r_\Gamma}
\newcommand\e{\epsilon}

\newcommand{\beq}{\begin{equation}}
\newcommand{\eeq}{\end{equation}}
\newcommand{\beqn}{\begin{eqnarray}}
\newcommand{\eeqn}{\end{eqnarray}}
\newcommand{\al}{\alpha}
\newcommand{\be}{\beta}
\newcommand{\cN}{{\cal N}}
\newcommand{\cM}{{\cal M}}
\newcommand{\cB}{{\cal B}}

\newcommand{\nn}{\nonumber}

\newcommand{\ro}{\rho}
\newcommand{\si}{\sigma}

\newcommand{\Mbar}{\overline{M}}
\newcommand{\gqqb}{gq\bar{q}}
\newcommand{\mZZ}{m_{ZZ}}
\newcommand{\pTcut}{p_{T,\mathrm{cut}}}
\newcommand{\sigmatree}{\sigma_I^{\mathrm{tree}}}
\newcommand{\mz}{m_Z}
\newcommand{\mw}{m_W}
\newcommand{\mv}{m_V}
\newcommand{\qgqqb}{qg+q\bar{q}}

\newcommand{\Do}{D_{\{1,3,2\}}}
\newcommand{\Dtw}{D_{\{2,1,3\}}}
\newcommand{\Dth}{D_{\{1,2,3\}}}
\newcommand{\Co}{C_{\{1,2\}}}
\newcommand{\Ctw}{C_{\{12,3\}}}
\newcommand{\Cth}{C_{\{1,3\}}}
\newcommand{\Cfo}{C_{\{2,3\}}}
\newcommand{\Cfi}{C_{\{1,23\}}}
\newcommand{\Bo}{B_{\{23\}}}
\newcommand{\Btw}{B_{\{123\}}}

\begin{document}

\title{Interference effects for Higgs-mediated $Z$-pair plus jet production}

\author{John M. Campbell}
\email{johnmc@fnal.gov}
\affiliation{Fermilab, PO Box 500, Batavia, IL 60510, USA}
\author{R. Keith Ellis}
\email{ellis@fnal.gov}
\affiliation{Fermilab, PO Box 500, Batavia, IL 60510, USA}
\author{ Elisabetta Furlan}
\email{efurlan@fnal.gov}
\affiliation{Fermilab, PO Box 500, Batavia, IL 60510, USA}
\author{Raoul R\"ontsch}
\email{rontsch@fnal.gov}
\affiliation{Fermilab, PO Box 500, Batavia, IL 60510, USA}
\date{\today}
\preprint{FERMILAB-PUB-14-315-T}

\begin{abstract}

We study interference effects in the production channel
$ZZ+$jet, in particular focusing on the role of the Higgs
boson. This production channel receives contributions both from Higgs
boson-mediated diagrams via the decay $H \to ZZ$ (signal
diagrams), as well as from diagrams where the $Z$-bosons couple 
directly to a quark loop (background diagrams).
We consider the partonic processes $gggZZ$ and $gq \bar{q} ZZ$ in which interference between
signal and background diagrams first occurs.
Since interference is primarily an off-resonant effect for the Higgs boson, we
treat the $Z$-bosons as on-shell. Thus our analysis is limited to the
region above threshold, where the invariant mass of the $Z$-pair,
$\mZZ$, satisfies the condition $\mZZ>2m_Z$.    In the region $m_{ZZ} >
300$~\GeV~we find that the interference in the $ZZ$~+~jet channel 
is qualitatively similar to interference in the inclusive $ZZ$~channel.
Moreover, the rates are sufficient to study these effects at the LHC once jet-binned data become available.

\end{abstract}

\maketitle

\section{Introduction}

The discovery of a Higgs boson at the Large Hadron Collider (LHC) with a mass of around $125$~GeV~\cite{Aad:2012tfa,Chatrchyan:2012ufa}
obliges us to undertake a program of precision measurements that will 
take more than a decade to complete. Since the observed particle, by virtue 
of its spin and the pattern of its couplings, is quite different than any other particle
observed to date, it will be important to examine {\it all} the features of this
particle. Kauer and Passarino made the interesting observation~\cite{Kauer:2012hd}
that the narrow width approximation is inadequate to describe the spectrum of the Higgs 
decay products in the channels $H \to VV$ (where $V$ is a vector boson). 
In fact a sizeable fraction $\sim 10\%$ of the Higgs-mediated cross section lies 
in a high mass tail, where the mass of the decay products is greater than $2 \mv$. 
The unique feature of this tail is that it is dependent on the couplings of the Higgs, 
both in production and decay, but, unlike measurements made on the Higgs boson peak, 
it is independent of the Higgs boson width. 
Subsequently a number 
of proposals have been made to exploit this high mass tail, 
either to bound the width of the Higgs boson~\cite{Caola:2013yja,Campbell:2013una,Campbell:2013wga} 
or to investigate the nature of the gluon-Higgs coupling~\cite{Cacciapaglia:2014rla,Azatov:2014jga}. 
Recent measurements by both the ATLAS~\cite{ATLAS-CONF-2014-042} and CMS~\cite{Khachatryan:2014iha} collaborations using this feature place 
bounds on the Higgs width at the level of $\Gamma_H / \Gamma_H^{\rm{SM}} \simeq 5-10$.
These bounds rely on the assumption that the 
Higgs couplings as measured off-shell agree with the Higgs couplings on-shell.
It is possible to construct models where this is not the case~\cite{Englert:2014aca}.
Independent of such measurements, however, an understanding of the off-shell behavior of the Higgs boson
is important in its own right and will be pursued vigorously once data-taking resumes at the LHC next year.

\begin{figure}
\includegraphics[scale=0.8,angle=0]{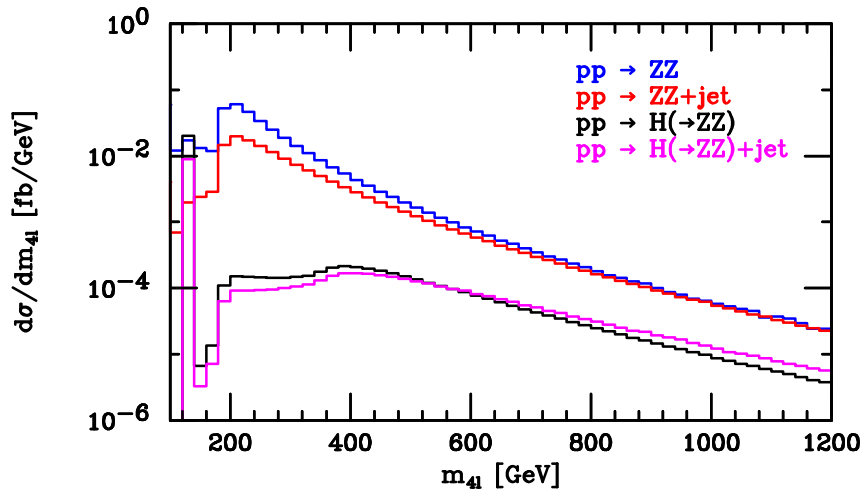}
\caption{Zero- and one-jet gluon fusion Higgs production with $H \to ZZ \to 4l$ decay as a function of the four-lepton invariant mass $m_{4l}$ (lower two curves), together with the leading order backgrounds $pp \to ZZ \to 4l$ and $pp \to ZZ (\to 4l) + \mathrm{jet}$ (upper two curves).
The results are for the LHC at $\sqrt{s}=8$ TeV, and a jet cut of $p_T>30$ GeV. The $Z$-bosons are taken to decay leptonically,
with a cut $m_{ll}>20$ GeV imposed on their decay products. All results are produced using MCFM~\cite{Campbell:2011bn}.} \label{fig:sigvslord}
\end{figure}

In this paper we consider $H \to ZZ$ decays, although
we expect qualitatively similar effects in the $H \to W^+W^-$
channel. The extraction of the Higgs contribution $gg \to H \to ZZ \to 4l$
in the high mass tail is challenging because the rate
is an order of magnitude smaller than the $pp \to ZZ \to 4l$
background, as shown in Fig.~\ref{fig:sigvslord}.
Since the gluon-gluon initial state
responsible for Higgs boson production radiates copiously, the cross
section for a Higgs boson produced in association with a jet is
expected to be large~\cite{Ellis:1987xu}.  
Indeed, as can be seen in Fig.~\ref{fig:sigvslord}, the production cross section for an off-shell Higgs with four-lepton invariant mass $m_{4l} > 300$~GeV
in the exclusive one-jet bin is comparable to that in the zero-jet bin, for a typical jet cut $p_{T,j}>30$~GeV.
On the other hand, for $m_{4l} > 300$ GeV, the cross section for the leading order (LO) exclusive $pp \to ZZ+\mathrm{jet}$
production is a factor of 1.5 smaller than that for inclusive $pp \to ZZ$.
It is thus clear that in
the tail, the ratio of Higgs signal to leading order background is better
in the one-jet bin than in the zero-jet bin.  
An examination of the gluon-Higgs coupling, or a constraint on the Higgs width,
from the one-jet bin can therefore be competitive with one from the zero-jet bin.
So far the ATLAS collaboration
has provided jet-binned differential distributions of the Higgs boson
only in the $\gamma \gamma$ channel~\cite{TheATLAScollaboration:2013eia}.  In
due course data will also become available in the $ZZ$ channel.  Anticipating this
development, in this paper we will study off-resonant Higgs boson effects in  the $ZZ+$jet channel.

For a calculation extending into the high mass tail, the dependence on the top mass must be retained.
The leading-order amplitudes contain one loop, and have been calculated in
Ref.~\cite{Ellis:1987xu}.  The next-to-leading order (NLO) corrections, which involve two-loop
amplitudes, are very challenging and have not yet been
calculated.   Note that, in the case where the heavy top limit is applicable, the relevant amplitudes have been known
to NLO for some time~\cite{Schmidt:1997wr,deFlorian:1999zd,Ravindran:2002dc,Glosser:2002gm},
and next-to-next-to-leading order (NNLO) results in the dominant $gg$ channel have been calculated
recently~\cite{Boughezal:2013uia,Chen:2014gva}.  

From the outset it was clear that an accurate description of the high mass region requires that interference
with non-Higgs mediated diagrams be taken into account~\cite{Glover:1988rg,Campbell:2011cu,Kauer:2012hd,Campanario:2012bh}. 
Indeed, the main role of the Higgs boson is to cancel the bad high-energy behavior that results from the
presence of longitudinal polarizations of $W$- and $Z$-bosons.
Thus the existence of this cancellation guarantees that interference
will be important in the high mass region. 
We must therefore also calculate non-Higgs mediated $ZZ+\mathrm{jet}$ production through a quark loop.

\renewcommand{\baselinestretch}{1.5}
\begin{table}[b]
\begin{tabular}{|c|l|c|}
\hline
Amplitude name& Process & Order of Amplitude \\
\hline
$\cM^{(a)}_3$ & $g + g \to H(\to Z Z) +g$       & $g_s^3 g_W^2$  \\
\hline
$\cB^{(a)}_3$ & $g + g \to Z Z +g$       & $g_s^3 g_W^2$  \\
\hline
\hline
$\cM^{(b)}_3$ & $q + g \to H(\to Z Z) +q$       & $g_s^3 g_W^2$  \\
\hline
$\cB^{(b)}_1$ & $q + g \to Z Z +q$       & $g_s g_W^2$  \\
\hline
$\cB^{(b)}_3$ & $q + g \to Z Z +q$       & $g_s^3 g_W^2$  \\
\hline
\hline
$\cB^{(c)}_1$ & $q + \bar{q} \to Z Z +g$ & $g_s g_W^2$  \\
\hline
\end{tabular}
\caption{Selection of parton processes for the $ZZ$~+ jet process. Representative diagrams
are shown in Figs.~\ref{gqqbZZ} and \ref{gggZZ}.
\label{gggZZtable}}
\end{table}
\renewcommand{\baselinestretch}{1}
\begin{figure}
\begin{center}
\includegraphics[angle=270,scale=0.70]{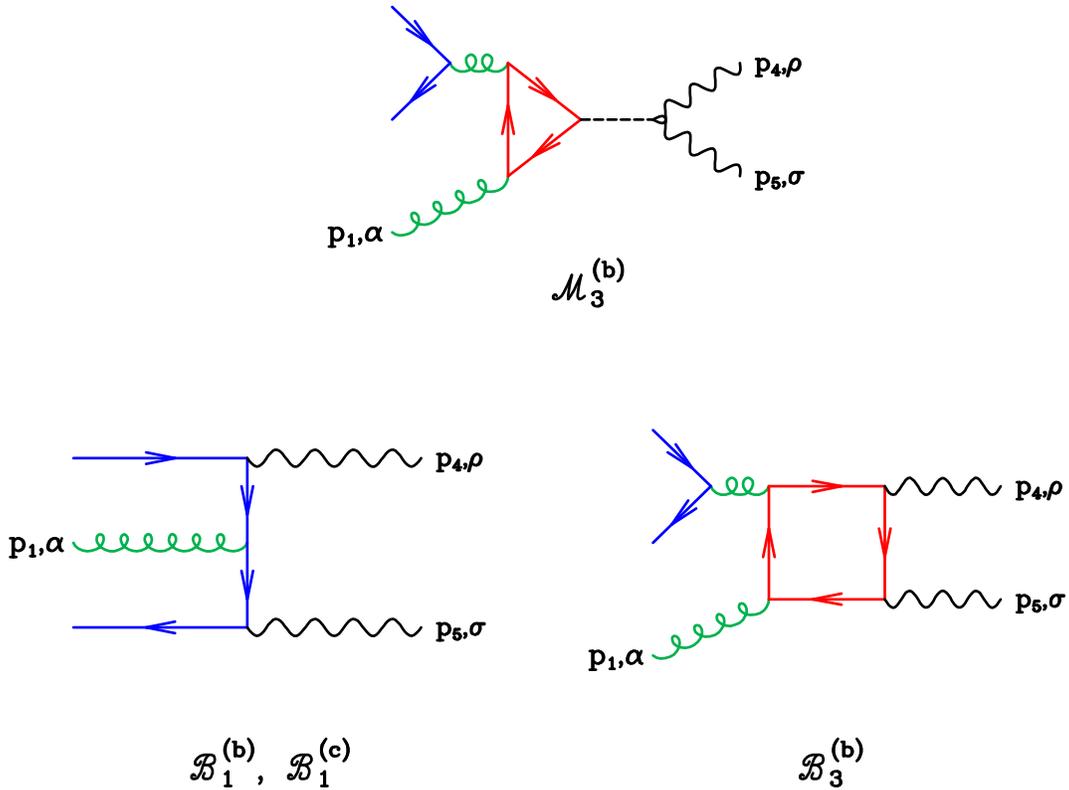}
\caption{Representative diagrams for the $0 \to gq \bar{q}ZZ$ amplitude.}
\label{gqqbZZ}
\end{center}
\end{figure}
\begin{figure}[b]
\begin{center}
\includegraphics[angle=270,scale=0.70]{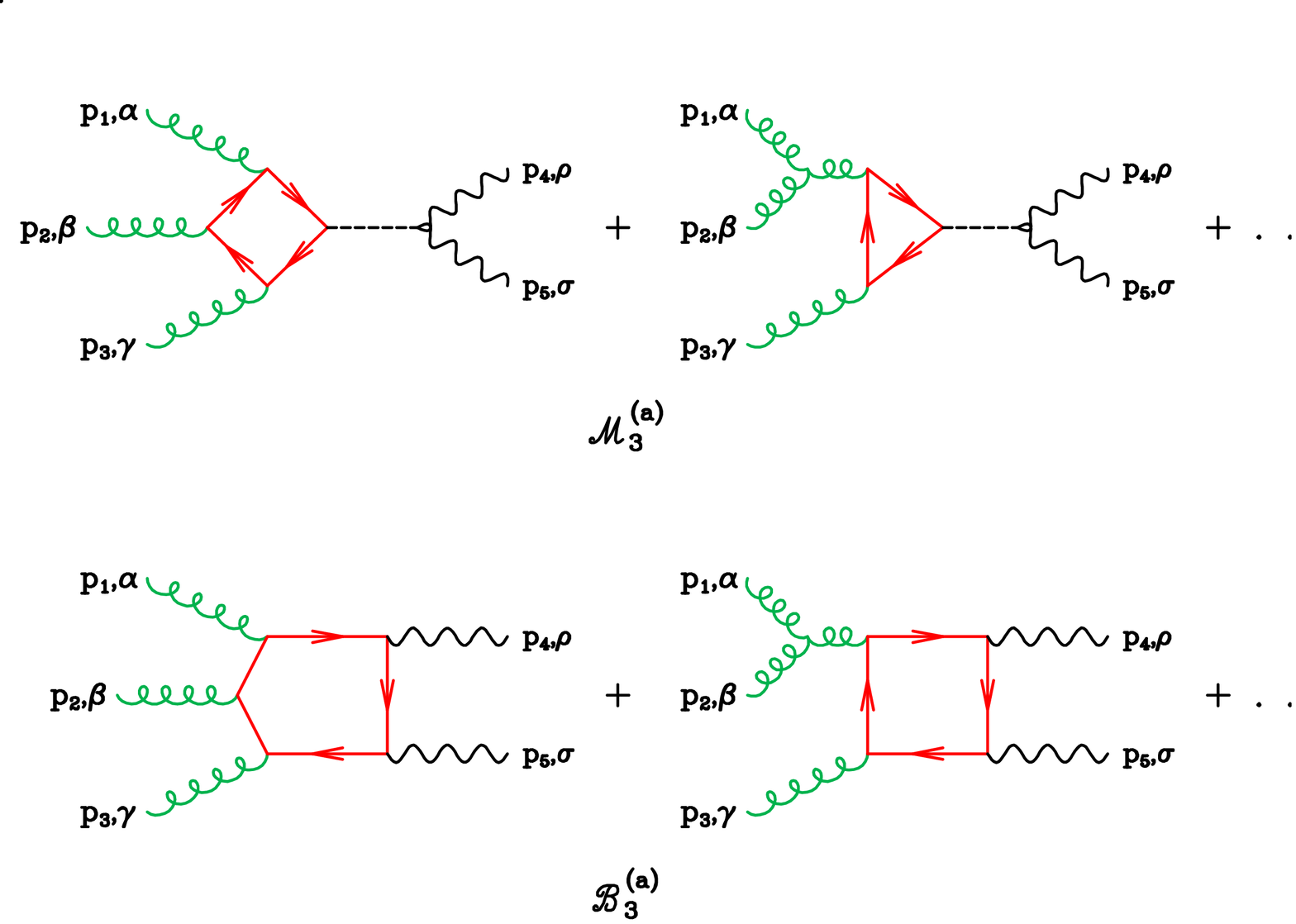}
\caption{Representative diagrams for the $0 \to gggZZ$ amplitude.}
\label{gggZZ}
\end{center}
\end{figure}
The relevant parton processes for the production of a pair of $Z$-bosons in association with a jet are given in Table~\ref{gggZZtable} and 
representative Feynman diagrams are shown in Figs.~\ref{gqqbZZ} and \ref{gggZZ}.
The lowest order at which this occurs is through the partonic reaction 
\beq \label{qqbar}
\cB^{(c)}_1:\;\; q + \bar{q} \to Z Z + g \; ,
\eeq
where the subscript indicates the order in the strong coupling, $g_s$, at which the amplitude first occurs. 
At the level of the matrix element squared, process~(\ref{qqbar}) enters at order $g_s^2 g_W^4$. 
The superscript is used to differentiate between partonic channels that enter at the same order.
For instance there are also the crossed processes with a gluon in the initial state, such as,
\beq
\cB^{(b)}_1:\;\; q + g \to Z Z + q \;. 
\label{qgtoq}
\eeq
The NLO corrections to the processes~(\ref{qqbar}) and~(\ref{qgtoq}), including all crossings, have been
presented in Refs.~\cite{Binoth:2009wk,Alwall:2014hca,Karg:2010wk}. 
In addition, ingredients for the NLO $ZZ$+jet process are part of the NNLO $ZZ$ calculation presented in 
Ref.~\cite{Cascioli:2014yka}. Merging to a parton shower generator has been considered in Ref.~\cite{Cascioli:2013gfa}.

Focusing on the one-loop corrections to the process~(\ref{qgtoq}), 
we find two classes of contributions that are separately gauge invariant and finite.
One represents box diagrams where the $Z$-bosons are radiated from a closed loop of fermions,
\beq
\cB^{(b)}_3:\;\; q + g  \xrightarrow{box} ZZ +q \;.
\label{qgtoZZq:box}
\eeq
and the other corresponds to diagrams in which a Higgs boson is produced through a massive quark loop
and subsequently decays to a pair of $Z$-bosons,
\beq 
\cM^{(b)}_3:\;\; q+g \to H(\to ZZ) +q\;.
\label{qgtoHq}
\eeq
The background process $\cB^{(b)}_3$ proceeds by a loop of quarks of all flavors, while $\cM^{(b)}_3$  receives significant contributions only for $t$ and $b$ quarks circulating in the loop.
Both processes~(\ref{qgtoZZq:box}) and~(\ref{qgtoHq}) interfere with the process~(\ref{qgtoq}), giving contributions of order $g_s^4 g_W^4$. 
However, the interference of (\ref{qgtoq}) and~(\ref{qgtoZZq:box}) is known to be small \cite{Binoth:2009wk}. 
Similarly, the interference between~(\ref{qgtoq}) and the Higgs-mediated process~(\ref{qgtoHq}) is also small in the inclusive case~\cite{Campbell:2013una}, as expected by unitarity.
We will verify in this paper that this hierarchy holds in the one-jet exclusive bin as well.
Partonic crossings of processes~(\ref{qgtoZZq:box}) and~(\ref{qgtoHq}) give rise to processes $ q + \bar{q}  \xrightarrow{box} ZZ +g$ and $q+\bar{q} \to H(\to ZZ) +g$ which interfere with process~(\ref{qqbar}).
However, these amplitudes can be trivially obtained from processes~(\ref{qgtoZZq:box}) and~(\ref{qgtoHq}), so there is no need to consider them separately.

At the next order, $g_s^6 g_W^4$, the squared loop amplitudes for production of a $Z$-pair in
association with a jet enter. As well as the square of the $qg$
processes~(\ref{qgtoZZq:box}) and~(\ref{qgtoHq}) and their interference,
gluon-induced production is also present at this order, either through Higgs production
\beq
\cM^{(a)}_3:\;\; g + g \to H(\to ZZ) +g\;,
\label{ggtoHg}
\eeq
or through loops of quarks in a similar fashion to the process~(\ref{qgtoZZq:box}),
\beq
\cB^{(a)}_3:\;\; g + g \to  Z Z +g \, .
\label{ggtoZZg}
\eeq

The gluon-induced process $\cB^{(a)}_3$ is known to provide contributions to $VV$ + jet production in the range of 5-10\%~\cite{Melia:2012zg,Agrawal:2012df,Cascioli:2013gfa}. 
Since these represent a NNLO correction to the continuum $pp \to ZZ$ + jet process we do not consider them in this work. 
Likewise, we do not consider the square of $\cB_3^{(b)}$. 
Instead, we confine our studies to the Higgs processes, i.e. the squares of processes~(\ref{qgtoHq}) and~(\ref{ggtoHg}), and interference between processes~(\ref{qgtoZZq:box}) and~(\ref{qgtoHq}), and between processes~(\ref{ggtoHg}) and~(\ref{ggtoZZg}).
The latter interference has been shown to lead to strong destructive interference in the high mass tail~\cite{Campanario:2012bh}.
Indeed, the results of Ref.~\cite{Campanario:2012bh} demonstrate the viability of studying interference effects in the Higgs + jet channel.
In this paper, we extend the analysis to include the interference of processes~(\ref{qgtoZZq:box}) and (\ref{qgtoHq}), which contributes at the level of 25-40\%, depending on the transverse momentum cut on the jet.

In addition to their use in studying interference effects in Higgs + jet production, the above contributions are necessary to extend the analysis of Ref.~\cite{Campbell:2013una,Campbell:2013wga} to NLO. 
This is especially important given the slow perturbative convergence of the Higgs cross sections and the corresponding large scale uncertainties 
and $k$-factors \cite{Dawson:1990zj,Djouadi:1991tka}.
In particular, the amplitudes of processes~(\ref{qgtoZZq:box})-(\ref{ggtoZZg}) make up the real radiation corrections to $gg \to ZZ$ and $gg \to H(\to ZZ)$.
For this reason, we obtain analytic formulae for the contributions arising from the amplitudes (\ref{qgtoZZq:box})-(\ref{ggtoZZg}), although the formulae are too
long to present in this paper.  This will allow a numerically stable computation of these amplitudes that can be integrated over the singular regions. 
This issue requires particular care because the real radiation amplitudes are at one-loop, rather than the tree-level amplitudes found in conventional NLO calculations.

The virtual contribution for the NLO calculation consists of two-loop amplitudes, including  the two loop $gg \to ZZ$ process. 
This is an extremely challenging calculation, which may be simplified if one limits
the scales involved to $s,t,m_q$ and $\mz$, where $s,t$ are the usual
Mandelstam variables and $m_q$ is the mass of the quark circulating in the
loop. 
We therefore consider both $Z$-bosons to be on their mass shells and sum over their
polarizations in this work, with a view to extending the calculation of gluon-induced inclusive $ZZ$ production to NLO.
Thus our calculation will be appropriate for the region
where the invariant mass of the $Z$-boson pair, $\mZZ > 2\mz$. 
\begin{figure}
\begin{center}
\includegraphics[angle=0,scale=0.7]{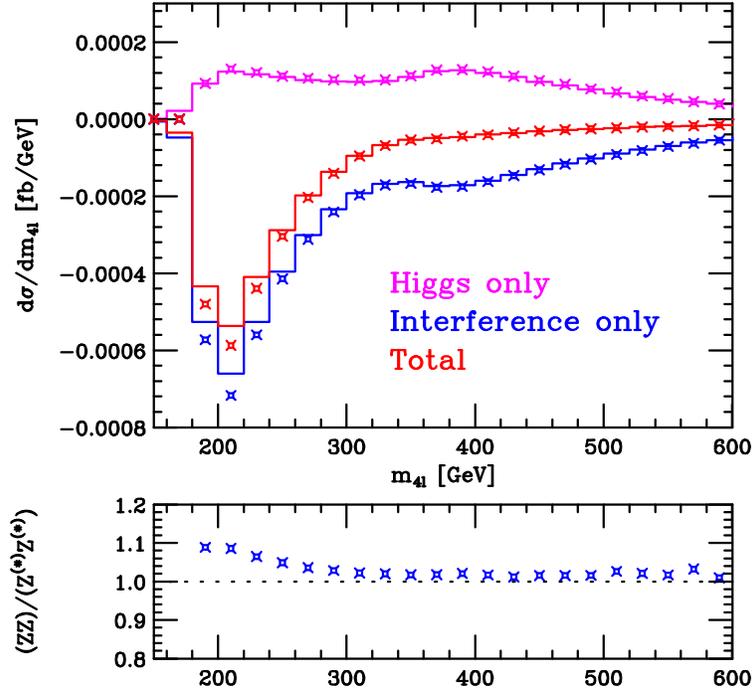}
\caption{Difference between the calculation of $gg \to Z^{(*)} Z^{(*)}$
(histograms), and $gg \to ZZ$ where both $Z$-bosons are on their
mass shells (points). The lower pane shows the ratio for the interference terms.
\label{fig:m4l_onshellcomparison}}
\end{center}
\end{figure}
As justification for this approximation, we can compare the calculation
of $gg \to Z^{(*)} Z^{(*)}$ from Ref.~\cite{Campbell:2013una}
with a simplified calculation where both $Z$-bosons are on their mass shells.
The results of this comparison are shown in Fig.~\ref{fig:m4l_onshellcomparison}.
Although there are differences between the calculations in the $ZZ$ threshold
region, for $m_{4l} > 300$~GeV the
two are essentially indistinguishable.

\section{Loop amplitudes for $ZZ$ production} \label{sec:ZZ}
\begin{figure}[ht]
\begin{center}
\includegraphics[angle=270,scale=0.7]{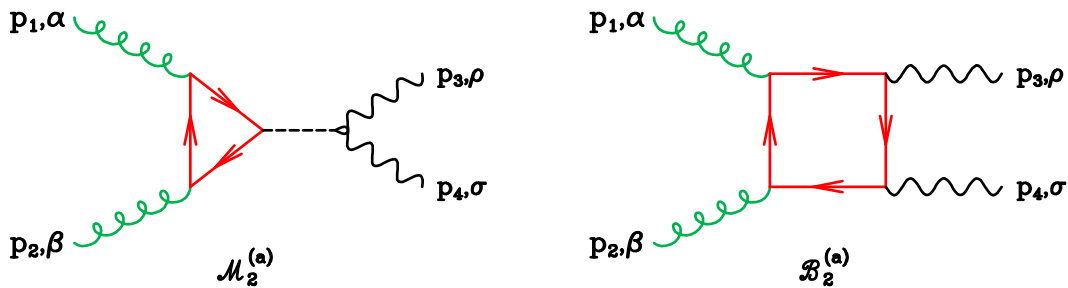}
\caption{Representative diagrams for the $0 \to ggZZ$ amplitude.}
\label{ggZZ}
\end{center}
\end{figure}

Although our principal focus will be the production of a pair of vector bosons in association with a jet,
we will first consider the process without a jet,  
which will serve to set up the notation.  Representative Feynman diagrams for this
process are shown in Fig.~\ref{ggZZ}. 
By analogy with processes~(\ref{ggtoHg}) and~(\ref{ggtoZZg}), we will refer to the amplitude for the Higgs signal process as $\cM_2^{(a)}$ and that for the continuum background as $\cB^{(a)}_2$, see Table~\ref{ggZZtable}. These amplitudes were first studied for on-shell $Z$-bosons in Ref.~\cite{Glover:1988rg}; more recently, the $Z$ decay and off-shell effects were also calculated \cite{Campbell:2013una}. 
In our results, we keep the $Z$-bosons on-shell and  sum over the polarizations.  
\renewcommand{\baselinestretch}{1.5}
\begin{table}[t]
\begin{tabular}{|c|l|c|}
\hline
Name & Process & Order of Amplitude \\
\hline
\hline
$\cM^{(a)}_2$ & $g + g \to H(\to Z Z) $       & $g_s^2 g_W^2$  \\
\hline
$\cB^{(a)}_2$ & $g + g \to Z Z$       & $g_s^2 g_W^2$  \\
\hline
\end{tabular}
\caption{Selection of parton processes for the $ZZ$ process. Representative diagrams are shown in Fig.~\ref{ggZZ}.
\label{ggZZtable}}
\end{table}
\renewcommand{\baselinestretch}{1}

\subsection{Amplitude for $gg \to H \to ZZ$}
We begin by looking at the amplitude for Higgs production, corresponding to the
left hand side of Fig.~\ref{ggZZ}(a), with all momenta outgoing.
The resultant amplitude is
\beq \cM^{(a), \alpha \beta}_{2}=-i \frac{g_W}{4 \mw}  \frac{g_s^2}{16 \pi^2} \frac{1}{2} \delta_{AB} \Mbar(s_H) \Bigg( g^{\alpha \beta} - \frac{p_2^{\alpha} p_1^{\beta}}{p_1\cdot p_2 }\Bigg)
\label{productionamplitude}
\eeq
where $A,B$ are the color indices,
$g_W=e/\sin\theta_W$ and $e,\mw,\theta_W$ are the electric charge, $W$-boson mass, and the Weinberg angle and $s_H=p_H^2 \equiv 2 p_1\cdot p_2$.
Our conventions for the Feynman rules are as given in Ref.~\cite{Ellis:1991qj}.
From Ref.~\cite{Georgi:1977gs}, the loop function for this amplitude is
\beq \label{A1}
\Mbar(s_H)= 8 m_q^2 \biggl[2- (s_H-4m_q^2) \Co \biggr]\, ,
\eeq
where $m_q$ is the mass of the quark circulating in the loop. $\Co$ is a scalar triangle integral; the exact definition is 
given in Table~\ref{integraldefns} and Appendix~\ref{Intdef}.
In the limit $m_q \to \infty$ 
we have that $\Mbar(s_H) \to \frac{8}{3}s_H$. 

\begin{table}
\begin{center}
\begin{tabular}{|l|l||l|l||}
\hline
$\Do$     &$D_0(p_1,p_{3},p_2;m,m,m,m)$&$\Co$&$C_0(p_1,p_2;m,m,m)$       \\
$\Dtw$     &$D_0(p_2,p_1,p_{3};m,m,m,m)$&$\Ctw$&$C_0(p_{12},p_{3};m,m,m)$ \\
$\Dth$    &$D_0(p_1,p_2,p_{3};m,m,m,m)$&$\Cth$&$C_0(p_{1},p_{3};m,m,m)$    \\
$\Bo$   &$B_0(p_{23};m,m)$   &$\Cfo$&$C_0(p_{2},p_{3};m,m,m)$ \\
$\Btw$  &$B_0(p_{123};m,m)$  &$\Cfi$&$C_0(p_{1},p_{23};m,m,m)$ \\
\hline
\end{tabular}
\end{center}
\caption{Definitions of the scalar integrals that appear in this paper. The notation for the scalar integrals follows Ref.~\cite{Passarino:1978jh}.}
\label{integraldefns}
\end{table}
For the calculation at hand we also need the decay amplitude, depicted on the right hand side of
Fig.~\ref{ggZZ}(a).  This amplitude is given by,
\beqn
\cM^{\rho \sigma}(H\to ZZ) &=&i g_W \frac{\mw}{\cos^2 \theta_W} g^{\rho \sigma} \; .
\label{decayamplitude}
\eeqn
Hence, combining Eqs.~(\ref{productionamplitude},\ref{decayamplitude}) the full amplitude for production and decay is
\beq
\cM^{(a),\alpha \beta \rho \sigma}_{2}  =    
 \cN \delta_{AB} \; \Mbar(s_H) \; \frac{1}{s_H-M_H^2} 
\Bigg( g^{\alpha \beta} -\frac{p_2^{\alpha} p_1^{\beta}}{p_1\cdot p_2}\Bigg) g^{\rho \sigma} ,
\eeq
where we have defined an overall normalization factor,
\beq
\cN=i \frac{g_W^2}{4 \cos^2 \theta_W} \frac{g_s^2}{16 \pi^2} \; \frac{1}{2} \; .
\label{eq:Ndef}
\eeq
From this it is straightforward to square the amplitude to obtain the result for the
Higgs diagrams alone.  The sum over the polarizations of the gluons and the $Z$-bosons of momentum $p$ can be
performed as usual with the projection operators,
\beq
P_g^{\mu \nu}=-g^{\mu \nu}, \hspace{1cm} P_Z^{\rho \beta}(p) = -g^{\rho \beta}+\frac{p^\rho p^\beta}{\mz^2} \, .
\label{Zpolsum}
\eeq
Including also the sum over colors yields the matrix element squared for the signal in this channel,
\beq
\mathcal{S}_{gg} \equiv\cM_{2}^{(a),\alpha\beta\rho'\sigma'} \ \bigl( \cM_{2,\alpha\beta\rho\sigma}^{(a)} \bigr)^* P_{Z \rho'}^{\rho}(p_3)P_{Z\sigma'}^{\sigma}(p_4)=|\cN|^2 \frac{V}{2} \frac{|\Mbar(s_H)|^2}{(s_H-M_H^2)^2}
 \bigg[ 8 + \bigg(\frac{s_{H}-2 \mz^2}{\mz^2}\bigg)^2\bigg]  \;,
\eeq
where we use the notation for the color factor $V=N_c^2-1=8$.

\subsection{Coupling structure for $gg \to ZZ$}
We turn now to the amplitude shown in Fig.~\ref{ggZZ}(b).
We are not interested in the square of this amplitude, which is an NNLO contribution to the continuum $pp \to ZZ$ production.
Rather, our focus is on the interference with the Higgs-mediated amplitude presented in the previous section.
We shall consider a single quark of flavor $f$ to be circulating in the quark loop.
The Standard Model coupling of this fermion to a $Z$-boson is given by,
\beq
-i \frac{g_W}{2 \cos\theta_W} \gamma^\mu (v_f - a_f \gamma_5),\;\;v_{f} = \tau_f -2 Q_f \sin^2 \theta_W,\;\;a_{f} = \tau_f,\;\;\tau_f =\pm \frac{1}{2}\, .
\eeq
The amplitude can be written by extracting an overall factor, given in Eq.~(\ref{eq:Ndef}),
\beq 
\cB_2^{(a),\al\be\ro\si}=\cN\; \delta_{AB} \Bigl[ v_f^2 B_{2,VV}^{(a),\al \be \ro \si}
+a_f^2 B_{2,AA}^{(a),\al \be \ro \si}
 + v_f a_f \left( B_{2,AV}^{(a),\al \be \ro \si}+B_{2,VA}^{(a),\al \be \ro \si} \right) \Bigr], \label{B2}
\eeq
where the $V$ and $A$ subscripts indicate the vector and axial vector coupling to the $Z$-bosons, respectively. The cross-terms proportional to $v_f a_f$ vanish, so that we can write
\beq 
\cB_2^{(a),\al\be\ro\si}=\cN\; \delta_{AB} \Bigl[ \left( v_f^2+a_f^2 \right) B_{2,VV}^{(a),\al \be \ro \si}
+a_f^2 \left( B_{2,AA}^{(a),\al \be \ro \si}-B_{2,VV}^{(a),\al \be \ro \si} \right) \Bigr]. \label{B2a}
\eeq
This decomposition of the coupling structure is particularly useful since the combination of amplitudes
$B_{2,AA}-B_{2,VV}$ vanishes in the limit $m_q \to 0$.

\subsection{Projection of interference for $gg \to ZZ$}

With the amplitudes outlined above it is straightforward to compute
the interference.  The relevant combination is,
\beqn
\mathcal{I}_{gg} &\equiv& 2 \left( \cM_{2,\al\be\rho'\sigma'}^{(a)} \right)^* {\cB}_{2}^{(a),\al\be\ro\si} P_{Z \rho}^{\rho'}(p_3)P_{Z\sigma}^{\si'}(p_4)\\
 &=& 2 V |\cN|^2 \; \Mbar(s_H)^* \; \frac{1}{s_H-M_H^2} 
\Bigg( g_{\al \be} -\frac{p_{2\al} p_{1\be}}{p_1\cdot p_2}\Bigg) \times \nn \\
& & \Bigl[ \left( v_f^2+a_f^2 \right) B_{2,VV}^{(a),\al \be \ro \si}
+a_f^2 \left( B_{2,AA}^{(a),\al \be \ro \si}-B_{2,VV}^{(a),\al \be \ro \si} \right) \Bigr] P_{Z\rho}^{\rho'}(p_{3})\; P_{Z\si\ro'}(p_{4}) \nn \\
 &=& 2 V |\cN|^2 \; \frac{2\, \Mbar(s_H)^*}{s_H-M_H^2} 
 \left[ (v_f^2+a_f^2) I_{VV} + a_f^2 \, I_{(AA-VV)} \right] \;,
\eeqn
where the projections are given by,
\beqn
I_{VV} &=& 32 \Bigg(       
       m_q^2 (2 p_1 \cdot p_2-\mz^2-2 m_q^2) (\Dth+\Dtw) \nn \\
      &+&(1-\frac{\mz^2+2 m_q^2}{p_1 \cdot p_2})     
        ((p_2 \cdot p_3 p_1 \cdot p_3+p_1 \cdot p_2 m_q^2-\frac{1}{2} p_1 \cdot p_2 \mz^2) \Do \nn  \\  
         &-& p_1 \cdot p_3 \Cth 
         -p_2 \cdot p_3 \Cfo)
       + 2 m_q^2 \Co+1\Bigg) \\
I_{(AA-VV)} &=& 64 m_q^2 \Bigg(  
         (\Do+\Dtw+\Dth) \frac{m_q^2}{\mz^4} (3 \mz^4-2 p_1 \cdot p_2 \mz^2+p_1 \cdot p_2^2) \nn\\ 
        &-& \frac{(p_1 \cdot p_2-3 \mz^2)}{\mz^2  p_1 \cdot p_2}\Big[                
        \frac{1}{2} (2 p_2 \cdot p_3 p_1 \cdot p_3-p_1 \cdot p_2 \mz^2) \Do- p_1 \cdot p_3 \Cth 
        -  p_2 \cdot p_3 \Cfo  \Big] \nn \\
       &-&p_1 \cdot p_2 (\Dth+\Dtw)
        +\Co \frac{p_1 \cdot p_2}{\mz^4} (p_1 \cdot p_2 - \mz^2) \Bigg) \;.
\eeqn

The notation for the scalar integrals $D$ and $C$ is given in Table~\ref{integraldefns}.
We note that, in contrast to the case where the $Z$-bosons are off-shell and their decays included,
these formulae for the interference take a very simple form.  In particular, there are no denominators
of the form $1/p_T^2$, where $p_T$ is the transverse momentum of one of the $Z$-bosons.

\section{Amplitudes for $ZZ+\rm{jet}$ production} \label{sec:ZZj}
We turn now to the amplitudes for $ZZ+\rm{jet}$ production. The partonic amplitudes are given in Eqs.~(\ref{qgtoq}-\ref{ggtoZZg}) and Table~\ref{gggZZtable} and are depicted in Figs.~\ref{gqqbZZ} and \ref{gggZZ}. The large-energy behavior of the background loop amplitudes $\cB_3^{(a)}$ and $\cB_3^{(b)}$ is unitarized by the Higgs amplitudes $\cM_3^{(a)}$ and $\cM_3^{(b)}$ respectively. 
In contrast, the amplitude $\cB_1^{(b)}$ is insensitive to the unitarizing effects of the Higgs boson. 

\subsection{Amplitude for $gq \to H(\to ZZ)q$} \label{sec:gqHq}
We begin by looking at the Higgs-mediated process~(\ref{qgtoHq}).
This was first computed in Ref.~\cite{Ellis:1987xu} for an on-shell Higgs. 
Modifying this result slightly to allow the Higgs to be off-shell, the amplitude is
\beq 
\cM^{(b),\alpha}_{3}=-i \frac{g_s^2}{16 \pi^2}  \; \frac{g_W}{4 \mw} \;  \frac{1}{2} (t^A)_{32} 
\; g_s \frac{1}{s_{23}} \bar{u}(p_3)\gamma_\mu u(p_2) 
\Bigg( g^{\alpha\mu}  - \frac{p_1^{\mu} (p_2^{\alpha}+p_3^{\al})}{p_1 \cdot (p_2+p_3)}\Bigg) F(s_{23},s_{H}) 
\eeq
where the loop function $F(s_{23},s_H)$ is given by
\beq
F(s_{23},s_{H})= -8 m_q^2 \left[2-(s_{H}-s_{23}-4 m_q^2)\Cfi + \frac{2 s_{23}}{s_{H}-s_{23}}\Big(\Btw-\Bo \Big)\right] \;,
\eeq
in terms of the scalar integrals defined in Table~\ref{integraldefns}.
Note that the loop function above is related to the $gg \to H$ loop function $\Mbar(s_H)$ given in Eq.~(\ref{A1}) by
\beq
F(0,s_H)=-\Mbar(s_H).
\eeq
Including the decay $H \to Z(p_4)Z(p_5)$, the amplitude is 
\beq \label{gqqbH}
\cM^{(b),\alpha \rho \sigma }_{3}=\cN
\Bigg(
g_s (t^A)_{32} \frac{F(s_{23},s_{H})}{s_{H}-M_H^2}\Bigg)
 \; \frac{1}{s_{23}}\bar{u}(p_3)\gamma_\mu u(p_2) 
\Bigg( g^{\al \mu}  - \frac{p_1^{\mu} (p_2^{\al}+p_3^{\al})}{p_1 \cdot (p_2+p_3)}\Bigg) 
\; g^{\rho \sigma}, 
\eeq
and squaring this we find 
\beqn
\mathcal{S}_{\gqqb} &\equiv& - \cM^{(b),\al \rho \sigma}_{3} \left( \cM_{3; \al \rho' \sigma' }^{(b)}\right)^*  P_{Z \rho}^{\rho'}(k_4)P_{Z\sigma}^{\sigma'}(k_5)\nn \\
&=&  \frac{V}{2} g_s^2 |{\cN}|^2 
\frac{1}{s_{23}}  \frac{\big( p_1\cdot p_{2}^2+ p_1\cdot p_3^2\big)}{p_1 \cdot p_{23}^2}
\frac{|F(s_{23},s_{H})|^2}{(s_{H}-M_H^2)^2}
\bigg[ 8 + \bigg(\frac{s_{H}-2 \mz^2}{\mz^2}\bigg)^2\bigg].
\eeqn
The negative sign in the first line comes from the sum over the gluon polarizations. Recall that $ {\cN}$ 
is our canonical overall factor given in Eq.~(\ref{eq:Ndef}).

\subsection{Amplitude for tree-level $qg\to ZZ q$}
The tree-level background amplitude in process~(\ref{qgtoq}) is given by
\beq
\begin{split}
\cB_{1}^{(b),\alpha \rho \sigma}=&\frac{i g_s g_W^2}{4\cos^2\theta_W}(t_A)_{32}\bar{u}(p_3) \bigl( (v_f^2+a_f^2)+2v_f a_f\gamma_5 \bigr) \times \\
&\bigl( T^{\alpha \rho \sigma}(p_1,p_2,p_3,p_4,p_5)+ T^{\alpha\sigma\rho}(p_1,p_2,p_3,p_5,p_4)\bigr) v(p_2),
\end{split}
\eeq
where the gamma-matrix structure is contained in the function $T^{\alpha \rho \sigma}$:
\beq
T^{\alpha \rho\sigma}(p_1,p_2,p_3,p_4,p_5)=\frac{\gamma^{\alpha}\hat{p}_{13}\gamma^{\rho}\hat{p}_{25}\gamma^{\sigma}}{s_{13}s_{25}}+\frac{\gamma^{\rho}\hat{p}_{34}\gamma^{\sigma}\hat{p}_{12}\gamma^{\al}}{s_{12}s_{34}}+\frac{\gamma^{\rho}\hat{p}_{34}\gamma^{\alpha}\hat{p}_{25}\gamma^{\sigma}}{s_{34}s_{25}}.
\eeq
This amplitude squared is the leading-order contribution to $pp \to ZZj$. Its interference with the Higgs-mediated amplitude, Eq.~(\ref{gqqbH}) is
\beq
\mathcal{I}_{\gqqb}^{(4)} = -2 \cM^{(b),\mu \rho \sigma }_{3} \left( \cB^{(b)}_{1,\mu \rho' \sigma'}\right)^*  P_{Z \rho}^{\rho'}(k_4)P_{Z\sigma}^{\sigma'}(k_5)
\eeq
where the superscript indicates that the interference is at order $g_s^4$. 

\subsection{Projection of interference for $qg \to ZZ q$}
The amplitude of process~(\ref{qgtoZZq:box}) has the same weak coupling structure as $gg \to ZZ$, and can therefore be written in terms of vector and axial couplings as in Eq.~(\ref{B2a}),
\beq
\cB_{3}^{(b),\al \rho \sigma}={\cN} g_s(t_A)_{32} \bar{u}(p_3) \gamma_{\nu} u(p_2)
\Bigl[ \left( v_f^2+a_f^2 \right) B_{3,VV}^{(b),\nu \al \ro \si}+a_f^2 \left( B_{3,AA}^{(b),\nu \al \ro \si}-B_{3,VV}^{(b),\nu \al \ro \si} \right) \Bigr] \;,
\eeq 
where $B^{(b)}_{3,VV}$ and $B^{(b)}_{3,AA}$ are loop functions.
Again, our focus is not on the square of this amplitude but on its interference with the Higgs-mediated process presented in Sec.~\ref{sec:gqHq}.
This is given by 
\beqn
\mathcal{I}^{(6)}_{\gqqb} &\equiv& -2 \cM^{(b)}_{3,\al \rho' \sigma'} \left( \cB_{3}^{(b),\al\rho\sigma}\right)^* \nn  P_{Z \rho}^{\rho'}(k_4) P_{Z \sigma}^{\sigma'}(k_5) \\
&=& -2 |\mathcal{N}|^2 g_s^2 C_F N_C \frac{F(s_{23},s_{H})}{s_{H}-M_H^2}\frac{1}{s_{23}}\mathrm{Tr} \bigl[\hat{p}_3\gamma^{\mu}\hat{p}_2\gamma_{\nu}\bigr]
\Bigg( g_{\al \mu}  - \frac{p_{1 \mu} (p_{2 \al}+p_{3 \al})}{p_1 \cdot (p_2+p_3)}\Bigg) \nn \\
&&\times   P_{Z \rho}^{\rho'}(k_4) P_{Z \rho' \sigma}(k_5) \Bigl[ \left( v_f^2+a_f^2 \right) B_{3,VV}^{\nu \al \ro \si}+a_f^2 \left( B_{3,AA}^{(a),\nu \al \ro \si}-B_{3,VV}^{(a),\nu \al \ro \si} \right) \Bigr].
\eeqn

\subsection{Amplitude for $gg \to H(\to ZZ) g $}
We now move on to the Higgs and interference contributions through gluon-fusion.
The Higgs-mediated contribution, represented by the diagrams
in Fig.~\ref{gggZZ}(a), is also presented in Ref.~\cite{Ellis:1987xu}.
By combining this with the decay amplitude given in
Eq.~(\ref{decayamplitude}), we obtain the full amplitude for the process
at hand, 
\begin{eqnarray}
&&\cM^{(a),\alpha \beta \gamma\rho \sigma }_{3}= -4  \cN g_s g^{\rho \sigma} \frac{s_H^2}{s_H-M_H^2} f_{ABC}   
\Bigg[ F_2^{\alpha \beta \gamma }(p_1,p_2,p_3) A_3(p_1,p_2,p_3)    \nonumber \\
      &+&   F_1^{\alpha \beta \gamma}(p_1,p_2,p_3) A_2(p_1,p_2,p_3)   
      +F_1^{\beta \gamma \alpha}(p_2,p_3,p_1) A_2(p_2,p_3,p_1) 
      +F_1^{\gamma \alpha \beta}(p_3,p_1,p_2) A_2(p_3,p_1,p_2) \Bigg]. \nn \\
\label{gggZZamp}
\end{eqnarray}
The projectors $F_1$ and $F_2$ are defined by,
\beqn
F_1^{\alpha \beta \gamma}(p_1,p_2,p_3)&=&
               \Bigg(\frac{g^{\alpha\beta}}{p_1\cdot p_2}-\frac{p_1^{\beta} p_2^{\alpha}}{p_1\cdot p_2^2}\Bigg) 
\Bigg(\frac{p_2^{\gamma}}{p_2\cdot p_3}-\frac{p_1^{\gamma}}{p_1\cdot p_3}\Bigg) \nn \\
F_2^{\alpha \beta \gamma}(p_1,p_2,p_3)&=&
              \frac{p_3^{\alpha} p_1^{\beta} p_2^{\gamma}-p_2^{\alpha} p_3^{\beta} p_1^{\gamma}}
{ p_1\cdot p_2 \, p_1\cdot p_3 \, p_2\cdot p_3}  +  \frac{g^{\alpha \beta}}{p_1\cdot p_2} \Bigg(\frac{p_1^{\gamma}}{p_3\cdot p_1}-\frac{p_2^{\gamma}}{p_3\cdot p_2}\Bigg)\nonumber \\
              &+&  \frac{g^{\beta \gamma}}{p_2\cdot p_3} \Bigg(\frac{p_2^{\alpha}}{p_1\cdot p_2}-\frac{p_3^{\alpha}}{p_1\cdot p_3}\Bigg)+  \frac{g^{\alpha\gamma}}{p_1\cdot p_3}  \Bigg(\frac{p_3^{\beta}}{p_2\cdot p_3}-\frac{p_1^{\beta}}{p_2\cdot p_1}\Bigg) 
\label{tensordefn}
\eeqn
and the functions $A_2$ and $A_3$ contain the loop integral functions, whose
definition we do not repeat here.  Instead, we note that the function $A_3$
is totally symmetric under the interchange of its arguments
while $A_2$ is symmetric only in its first two arguments,
\beq
A_2(p_1,p_2, p_3) = A_2(p_2,p_1,p_3)\, .
\eeq
The amplitude-squared for this contribution, summed over colors and spins, is
then given by,
\beqn \label{Sggg}
\mathcal{S}_{ggg} &\equiv& - \cM^{(a),\alpha \beta \gamma\rho \sigma }_{3} \left( \cM_{3,\alpha \beta \gamma \rho' \si' }^{(a)} \right)^* P_{Z \rho}^{\rho'}(k_4)P_{Z\sigma}^{\si'}(k_5) \nn \\
&=&|\cN|^2 g_s^2 \frac{64 V N}{s_{12} s_{23} s_{31}} \frac{s_H^4}{(s_H-M_H^2)^2}
\Big[8+\Big(\frac{s_H-2\mz^2}{\mz^2}\Big)^2\Big] \nn \\
&\times & \Big[|A_2(p_1,p_2,p_3)|^2+|A_2(p_2,p_3,p_1)|^2
 +|A_2(p_3,p_1,p_2)|^2+|A_4(p_1,p_2,p_3)|^2\Big] \;.
\eeqn
The negative sign in the first line comes from the sum over the polarizations of the three gluons. We have introduced a new function $A_4$ that is defined by,
\beq
A_4(p_1,p_2,p_3)= [ A_2(p_1,p_2,p_3) + A_2(p_2,p_3,p_1) + A_2(p_3,p_1,p_2) - 2 A_3(p_1,p_2,p_3 ) ] \;. 
\eeq

As a cross-check, we can inspect the limit $m_q \to \infty$ in which these 
functions take the limiting values,
\beq
A_4(p_1,p_2,p_3)=-\frac{1}{3}; \hspace{0.5cm} A_2(p_1,p_2,p_3)=-\frac{4 p_1 \cdot p_2^2}{3 s_H^2}.
\eeq
Thus in this limit the squared amplitude becomes,
\beq
\mathcal{S}_{ggg} =|\cN|^2 g_s^2 \frac{64 V N}{9} \frac{1}{(s_H-M_H^2)^2}
\left(8+\Big(\frac{s_H-2\mz^2}{\mz^2}\Big)^2\right)  \left( \frac{s_H^4 + s_{12}^4 + s_{23}^4 + s_{31}^4}{s_{12} s_{23} s_{31}} \right) ,\;
\eeq
which is the expected result~\cite{Dawson:1990zj}.

\subsection{Projection of interference for $gg \to ZZ g$}
Finally, we turn to the background amplitude of process~(\ref{ggtoZZg}). 
The relevant topologies of diagrams are 
shown in Fig.~\ref{gggZZ}(c,d). There are 42 diagrams in all, 24 of the topology of Fig.~\ref{gggZZ}(c) 
and 18 of the topology of Fig.~\ref{gggZZ}(d). 
The continuum amplitude can be written as,
\beq
\cB_{3}^{(a),\alpha\beta\gamma\ro\sigma} =-2 i g_s\cN \left[ {\rm Tr}\left(T^{A} T^{B} T^{C}\right) B_3^{(a),\alpha\beta\gamma\ro\sigma}(1,2,3)+{\rm Tr}\left(T^{A} T^{C} T^{B}\right) B_3^{(a),\alpha\gamma\beta\ro\sigma}(1,3,2) \right]
\eeq
where $ B_3^{(a),\alpha\beta\gamma\ro\si}(1,2,3)$ and  $B_3^{(a),\alpha\gamma\beta\ro\si}(1,3,2)$ are color-ordered gauge-invariant primitive amplitudes. 
Since we are only interested in the interference between $\cB_3^{(a)}$ and the Higgs amplitude $\cM_3^{(a)}$, and since  $\cM_3^{(a)}$ is proportional to the antisymmetric color structure $f^{ABC}$ (cf. Eq.~(\ref{gggZZamp})), terms proportional to the symmetric color combination $d^{ABC}$ will vanish in the interference and can be safely dropped.
Thus the amplitude can be replaced by,
\beq
\cB_{3}^{(a),\alpha\beta\gamma\rho\sigma} =  \frac{1}{2}g_s\cN   f^{ABC} \left[
 B_3^{(a),\alpha\beta\gamma\rho\sigma}(1,2,3) - B_3^{(a),\beta\alpha\gamma\rho\sigma}(2,1,3) \right].
\eeq
The weak coupling structure of the amplitudes $B_3^{(a)}$ can be written as a linear combination of $v_f^2$, $a_f^2$ and $v_fa_f$, cf. Eq.~(\ref{B2}).
By inspection of the diagrams it can be seen that, for either purely vector ($VV$) or purely
axial ($AA$) couplings in the loop, $B_3^{(a),\beta\alpha\gamma\rho\sigma}(2,1,3) = -B_3^{(a)\alpha\beta\gamma\rho\sigma}(1,2,3)$.
In contrast, for the mixed case ($VA$ or $AV$) the two permutations are equal,
$B_3^{(a),\beta\alpha\gamma\rho\sigma}(2,1,3) = B_3^{(a),\alpha\beta\gamma\rho\sigma}(1,2,3)$.  Hence we can simply write,
\beq
\cB_{3}^{(a),\alpha\beta\gamma\rho\sigma} = g_s\cN f^{ABC} B_3^{(a),\alpha\beta\gamma\rho\sigma}(1,2,3)
\eeq
and only consider two combinations of vector boson couplings, $VV$ and $AA-VV$.

The interference is given by
\beq
\mathcal{I}_{ggg} = - 2 \cM^{(a),\alpha\beta\gamma\ro'\sigma'}_3 \left( \cB_{3,\alpha \beta \gamma \rho \si }^{(a)} \right)^* P_{Z \rho'}^{\rho}(k_4)P_{Z\sigma'}^{\si}(k_5),
\eeq
where again the minus sign comes from the sum over the gluon polarizations. 
Our strategy will be to contract the continuum amplitudes with the tensors $F_1$ and $F_2$ present in the Higgs amplitude, Eq.~(\ref{gggZZamp}). The definitions of the tensors is given in Eq.~(\ref{tensordefn}).
Writing this explicitly, we have
\beqn
\mathcal{I}_{ggg}  &=&8|\cN|^2 g_s^2 VN\frac{s_H^2}{(s_H-M_H^2)} \, P_{Z \rho'}^{\rho}(k_4)P_{Z}^{\rho'\sigma}(k_5) \nn \\
&& \Biggl[  \biggl(A_3(1,2,3)F_2^{\alpha\beta\gamma}(1,2,3) + \sum_{\tilde{P}(1,2,3)} A_2(1,2,3)  F_1^{\alpha\beta\gamma}(1,2,3) \biggr) \Biggr] \left( B^{(a)}_{3,\alpha\beta\gamma\rho\sigma}(1,2,3) \right)^*
\eeqn
where the sum is over the three cyclic permutations. Since $B_3^{(a)}$ is fully symmetric under such permutations, we can write the above as
\beq
\mathcal{I}_{ggg} = 8 |\cN|^2 g_s^2 VN\frac{s_H^2}{(s_H-M_H^2)} 
 \Biggl[A_3(1,2,3) H_3(1,2,3) + \sum_{\tilde{P}(1,2,3)} A_2(1,2,3)H_2(1,2,3) \Biggr]
\eeq
where
\beqn
H_3(1,2,3)&=&F_2^{\alpha\beta\gamma}(1,2,3)\left(B^{(a)}_{3,\alpha\beta\gamma\rho\sigma}(1,2,3) \right)^* \, P_{Z \rho'}^{\rho}(k_4)P_{Z}^{\rho'\sigma}(k_5)\nn \\
H_2(1,2,3)&=&F_1^{\alpha\beta\gamma}(1,2,3)\left(B^{(a)}_{3,\alpha\beta\gamma\rho\sigma}(1,2,3) \right)^* \, P_{Z \rho'}^{\rho}(k_4)P_{Z}^{\rho'\sigma}(k_5).
\eeqn
In our implementation we have analytically computed $H_3(1,2,3)$ and $H_2(1,2,3)$
and then performed the sum over the three permutations numerically.

\section{Results} \label{sec:results}
The Higgs and interference amplitudes presented in Secs.~\ref{sec:ZZ} and~\ref{sec:ZZj} have been implemented in the parton level
integrator MCFM, using a library of scalar integrals~\cite{vanHameren:2010cp}.
In this section we present results for the LHC running at $\sqrt{s}=8$ TeV and $\sqrt{s}=13$ TeV.
Our parameters are summarized in Table~\ref{tab:params}. 
Since we are particularly interested in the behavior of the high mass tail, we make use of a dynamic factorization/renormalization scale $\mu=m_{ZZ}/2$.
We remind the reader that we consider on-shell $Z$-bosons, and include their decay only through a branching ratio $BR(Z \to e^+e^-)=3.36386 \times 10^{-2}$. 
Thus, we are insensitive to the details of the lepton kinematics.
We demand the presence of a single jet, defined using the anti-$k_T$ algorithm and having a rapidity $|\eta_j|<3$ and a transverse momentum $p_{T,j} > \pTcut$.
We make use of the MSTW08LO parton distributions functions (pdfs) throughout~\cite{Martin:2009iq}.

\begin{table}
\begin{center}
\begin{tabular}{|ll|ll|ll|}
\hline
$m_H$&$=126~\mathrm{GeV}$ & $\Gamma_H$&$=4.307~\mathrm{MeV}$ & $m_Z$ &$= 91.1876~\mathrm{GeV}$ \\
$m_t$&$=173.225~\mathrm{GeV}$ & $m_b$&$=4.75~\mathrm{GeV}$ & $\sin^2 \theta_W$ &$= 0.2226459$ \\
$G_F$&$= 1.16639\times 10^{-5}~\mathrm{GeV}^{-2}$ & $g_W^2$ &$= 0.4264904$ &$e^2$ &$=0.0949563$ \\
\hline
\end{tabular}
\end{center}
\caption{Masses, widths and electroweak coupling parameters used in this work.} \label{tab:params}
\end{table}
We will refer to the cross sections that arise from the signal amplitudes $\cM^{(a)}_3$ and $\cM^{(b)}_3$ as $\sigma_{H}^{gg}$ and $\sigma_{H}^{\qgqqb}$ respectively, and their sum as $\sigma_H$. Similarly, the cross sections arising from the interference of these amplitudes with $\cB^{(a)}_3$ and $\cB^{(b)}_3$, respectively, are $\sigma_{I}^{gg}$ and $\sigma_{I}^{\qgqqb}$; their sum is $\sigma_I$.
The $q\bar{q}$-initiated contributions in $\sigma_{H}^{\qgqqb}$ and $\sigma_{I}^{\qgqqb}$ are at the level of less than one per-mille.
The cross section arising from the interference of  $\cM^{(b)}_3$  with the tree-level amplitude $\cB^{(b)}_1$ is $\sigmatree$.
Recall that we do not consider the square of background amplitudes such as $\cB_3^{(a)}$, as these contribute at NNLO to the continuum background.

In Table~\ref{tab:breakdown}, we show partonic level cross sections in the high mass tail defined by $m_{ZZ} >  300$~GeV. 
Four  different values of the jet cut $\pTcut$ are shown at center-of-mass energies $\sqrt{s}=8$ TeV and $\sqrt{s}=13$~TeV. 
We have confirmed that our results for $\sigma_H^{gg}$ and $\sigma_I^{gg}$ agree with those shown in Ref.~\cite{Campanario:2012bh}.
Comparing these $gg$-initiated cross sections with the $\qgqqb$-initiated cross sections (which are not considered in Ref.~\cite{Campanario:2012bh}), we see that the former are always larger but the latter are still important, especially at larger values of $\pTcut$. 
The relative partonic contributions at a given $\pTcut$ are roughly the same for Higgs and interference cross sections: at $\sqrt{s}=8$ TeV, 
both  $\sigma^{\qgqqb}_H / \sigma_H $ and $\sigma^{\qgqqb}_I /\sigma_I$ are approximately 25\% at $\pTcut=30$ GeV and increase to almost 50\% at $\pTcut=200$ GeV.
This effect is due to the larger $\pTcut$ probing a higher region of $x$, where the quark pdfs are relatively more important than the gluon pdfs.
At $\sqrt{s}=13$ TeV, the value of $x$ decreases, leading to smaller values for these ratios for a given $\pTcut$.

The negative values of the interference cross sections are required to restore unitarity.
These cross sections are slightly larger in magnitude than the
signal rate, so that their sum is negative.
The ratio of Higgs to interference is roughly constant for different values of $\sqrt{s}$ or $\pTcut$, and in either partonic channel.
In contrast, the tree-level interference
$\sigmatree$ is positive but fairly small, although its importance increases with $\pTcut$, for the reasons discussed above.

\begin{table}
\begin{center}
\begin{tabular}{|c|c|c|c|c|c|c|}
\hline
&$\pTcut$ [GeV] & $\sigma_H^{gg}$ [fb] & $\sigma_H^{\qgqqb}$[fb] & $\sigma_I^{gg}$[fb] & $\sigma_I^{\qgqqb}$[fb] & $\sigmatree$[fb] \\
\hline
\multirow{4}{*}{$\sqrt{s}=8$ TeV} & 30              &0.0212              &0.00679              & -0.0299           & -0.00929           & 0.00230    \\
                                  & 50              &0.0124              &0.00522              & -0.0173           & -0.00706           & 0.00182    \\
                                  & 100             &0.00467             &0.00279              & -0.00632          & -0.00369           & 0.00097   \\
                                  & 200             &0.00104             &0.00086             & -0.00133          & -0.00111           & 0.00026   \\

\hline
\multirow{4}{*}{$\sqrt{s}=13$ TeV} & 30             &0.0887             &0.0216              & -0.1263        & -0.0298           & 0.00652         \\
                                  & 50              &0.0547             &0.0172              & -0.0770        & -0.0235           & 0.00528         \\
                                  & 100             &0.0229            &0.0101              & -0.0313        & -0.0136           & 0.00298         \\
                                  & 200             &0.00612            &0.00377             & -0.00798       & -0.00497          & 0.00092         \\
\hline
\end{tabular}
\end{center}
\caption{Higgs and interference cross sections $\sigma_H$ and $\sigma_I$ by partonic channel, for four choices of $\pTcut$, at the
$\sqrt{s}=8$ TeV and $\sqrt{s}=13$ TeV LHC. All results shown are in the tail region $m_{ZZ}> 300$ GeV.}
\label{tab:breakdown}
\end{table}

In Table~\ref{tab:offshell_pTcut} we show $\sigma_H$ and $\sigma_I$ for the four values of $\pTcut$ and the two center-of-mass energies.
Also shown are the Higgs boson on-peak cross sections ($\sigma_H$), defined by $m_{ZZ} < 130$ GeV. 
The latter are obtained by a separate calculation in which the $Z$-bosons are allowed to be off-shell.
The cross sections at $\sqrt{s}=13$ TeV are a factor of 4--5 times larger than at $\sqrt{s}=8$ TeV, with greater increases coming from higher
values of $\pTcut$.
These values indicate that a few events have already been produced in the high mass tail of the one-jet exclusive bin during run I,
and about $100$ high mass events are expected with 300~$\mathrm{fb}^{-1}$ at the higher energy. 
We note that the high mass tail becomes more important relative to the peak cross section as
$\pTcut$ increases because the on-peak cross section is 
more concentrated at small transverse momentum.

As mentioned in the discussion of Table~\ref{tab:breakdown}, the sum of $\sigma_H$ and $\sigma_I$ is negative, and the ratio 
$|\sigma_I/\sigma_H|$ does not depend appreciably on $\pTcut$ or $\sqrt{s}$.
The value of $\sigmatree$ is about 5\% of the value of $\sigma_I$ with $\pTcut=30$ GeV, and around 10\% with $\pTcut=200$ GeV.
However this contribution will be partially cancelled by the unitarizing effect of the
interference between the tree-level amplitudes in process~(\ref{qgtoq}) and the one-loop box $qg$
process~(\ref{qgtoZZq:box}), which is neglected in this paper.
We therefore expect the values of $\sigma_{I,\mathrm{tail}}^{tree}$ given in Table~\ref{tab:offshell_pTcut}
to provide an upper bound on the size of the subleading contribution resulting from interference
effects involving the tree-level processes.

\begin{table}
\begin{center}
\begin{tabular}{|c|c|c|c|c|c|}
\hline
& $\pTcut$ [GeV] & $\sigma_{H,\mathrm{peak}}$ [fb] &
  $\sigma_{H,\mathrm{tail}}$ [fb] & $\sigma_{I,\mathrm{tail}}$ [fb] &
  $\sigma_{I,\mathrm{tail}}^{tree}$ [fb] \\
\hline
\multirow{4}{*} {$\sqrt{s}=8$ TeV} &  $30$  & 0.351  & 0.0280  & -0.0392  & 0.0023 \\
& $50$  & 0.206  & 0.0176  & -0.0244  & 0.0018 \\
& $100$ & 0.0714 & 0.0075 & -0.0100  & 0.0010 \\
& $200$ & 0.0128 & 0.0019 & -0.0024 & 0.00026 \\
\hline
\multirow{4}{*}{$\sqrt{s}=13$ TeV} & $30$  & 0.909 & 0.110  & -0.156 & 0.0065 \\
& $50$  & 0.557 & 0.0718 & -0.100  & 0.0053 \\
& $100$ & 0.212 & 0.0329 & -0.0448 & 0.0030 \\
& $200$ & 0.045 & 0.0099 & -0.0130 & 0.0009 \\
\hline
\end{tabular}
\end{center}
\caption{cross sections at $\sqrt{s}=8$ and $\sqrt{s}=13$ TeV in the peak region ($m_{ZZ}<130$~GeV) and
in the high mass tail region defined by $m_{ZZ}>300$~GeV,
for $\sigma_H$ and $\sigma_I$.
Also shown is the tree-level interference $\sigma_I^{tree}$.}
\label{tab:offshell_pTcut}
\end{table}

\begin{figure}
\begin{center}
\includegraphics[angle=0,scale=0.8,angle=0]{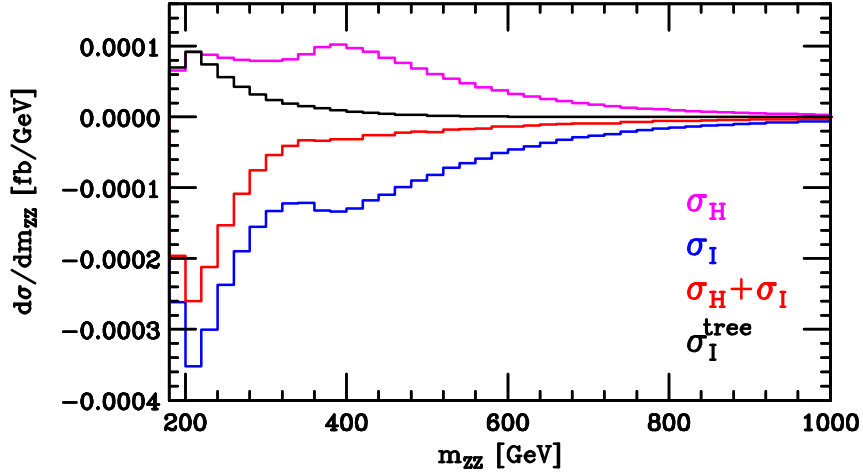}
\caption{Higgs and interference distributions $\sigma_H$ and $\sigma_I$, and their sum,
for the invariant mass of the $Z$- pair, $m_{ZZ}$. Also shown are the tree-level interference distributions $\sigmatree$.
The results were obtained using $\pTcut = 30$ GeV and $\sqrt{s}=8$~TeV.
\label{fig:zzj_m4l}}
\end{center}
\end{figure}

The dependence of these quantities on the invariant mass of the $Z$-pair $m_{ZZ}$ is shown  in Fig.~\ref{fig:zzj_m4l} for $\pTcut=30$ GeV and $\sqrt{s}=8$ TeV.
We see that the inclusion of the interference term changes the sign of the Higgs-mediated contribution and its magnitude is
significantly reduced, as anticipated.  Moreover, the effect of the interference is to dramatically alter
the shape of the distribution throughout.
This emphasizes the importance of including the interference effects when considering the high mass tail.

As of yet, no mass distributions are available in the one-jet bin, so an extraction of a bound on the Higgs width using our results is not possible. 
As data become available from the higher energy LHC run, this analysis will become possible.
We expect that such a bound will be competitive with the bound extracted from the zero-jet bin. 
To see this, we can examine the cross section for Higgs-mediated events in the high mass region.
We assume that the on-shell Higgs cross-section corresponds to its SM value. This introduces a relationship between the couplings and the Higgs width, which can be used to write the high mass cross-section in terms of the width~\cite{Caola:2013yja,Campbell:2013una}
\beq
\sigma_{off,ZZ+jet}^{H+I}(m_{ZZ} > 300~\mathrm{GeV})=\sigma_H(m_{ZZ} > 300~\mathrm{GeV}) \left( \frac{\Gamma_H}{\Gamma_H^{SM}} \right)
 + \sigma_I(m_{ZZ} > 300~\mathrm{GeV}) \sqrt{\frac{\Gamma_H}{\Gamma_H^{SM}}} \;.
\eeq
The result for the $4$-lepton final state presented in Ref.~\cite{Campbell:2013una} was,
\begin{equation}  \label{ZZoffshell}
\sigma_{off, 4\ell{\rm -fiducial}}^{H+I}(m_{4\ell}>300~\rm{GeV}) 
 = 0.025 \left( \frac{\Gamma_H}{\Gamma_H^{SM}} \right) - 0.036 \sqrt{\frac{\Gamma_H}{\Gamma_H^{SM}}}~{\rm fb},
\end{equation}
where the $Z$-bosons were produced off-shell, and fiducial cuts were applied to the leptons originating from their decay.
In order to compare with the methodology of this paper, we repeat the calculation of Ref.~\cite{Campbell:2013una} but keep the $Z$-bosons on-shell and simply apply
the appropriate branching ratio into leptons. In this case the result is,
\begin{equation} \label{ZZonshell}
\sigma_{off,ZZ}^{H+I}(m_{ZZ}>300~\rm{GeV}) 
 = 0.0323 \left( \frac{\Gamma_H}{\Gamma_H^{SM}} \right) - 0.0468 \sqrt{\frac{\Gamma_H}{\Gamma_H^{SM}}}~{\rm fb}.
\end{equation}
The relative size of the two coefficients in Eqs.~(\ref{ZZoffshell}) and (\ref{ZZonshell}) is the same, as might be expected.
In the presence of an additional jet, defined with a $p_T$ cut of $30$~GeV, the
equivalent result is:
\begin{equation}
\sigma_{off,ZZ+jet}^{H+I}(m_{ZZ}>300~\rm{GeV}) 
 = 0.0280 \left( \frac{\Gamma_H}{\Gamma_H^{SM}} \right) - 0.0392 \sqrt{\frac{\Gamma_H}{\Gamma_H^{SM}}}~{\rm fb},
\end{equation}
where the coefficients have been read from Table~\ref{tab:offshell_pTcut}.
Thus the prediction for the effect of the Standard Model Higgs boson on the
number of off-shell $ZZ$+jet events is slightly
smaller than the effect on the number of off-shell $ZZ$ events,
inclusive in the number of jets.
However the scaling with a non-SM value of the width is about the same.
The equivalent formulae for other values of the jet cut can be read off from
Table~\ref{tab:offshell_pTcut}.
We anticipate a more detailed study once further experimental data on $ZZ$ production in the one-jet bin are available.

\section{Conclusions} \label{sec:concl}
We have studied the high mass tail of a Higgs boson produced in association with one jet, focusing on the Higgs decay to a $Z$-boson pair.
We have performed the calculation in a simple kinematic configuration with on-shell $Z$-bosons which are summed over polarizations.
The overall Higgs rate in the one-jet bin is known to be large, and we find a significant contribution from the high mass tail
for a typical jet transverse momentum and rapidity.  This feature has already been noted in the inclusive case, where
it has been used to extract a tight bound on the Higgs width.
In addition, the ratio of the Higgs signal to the dominant leading order background $pp \to ZZ + n$ jets is larger in the one-jet bin than in the zero-jet bin.
It is therefore desirable to study the high mass tail in the one-jet bin, both in current and future LHC data.

An accurate prediction of the high mass tail requires an understanding of the interference between Higgs-mediated and non-Higgs mediated amplitudes, and we have studied this interference from both the $gg$ and $qg$ production modes. 
Qualitatively, the effects of the interference are similar to those found in the inclusive case: the interference provides a negative contribution to the high mass tail which is larger in magnitude than the signal rate, leading to a negative shift in the distributions. 
The interference between the Higgs-mediated one-loop diagrams and tree-level background diagrams is found to be subdominant, despite these entering at a lower order in $g_s$.
A brief analysis shows that the bounds on the Higgs width that can be extracted in this channel are comparable to those from inclusive production.
We also point out that the results presented here form an important step to extending the understanding of the high mass tail in the inclusive case to NLO. 

Throughout this paper we have considered Higgs production through gluon fusion and neglected
the subdominant production mechanism of weak boson fusion (WBF). 
This mechanism has a distinctive signature of two very forward jets, with little hadronic activity between them.
These signatures dominate the two-jet bin, but will also contribute to the one-jet bin if one of the jets is missed by the detector.
A Higgs boson produced through WBF must necessarily exhibit similar high mass tail effects, if the Higgs is to unitarize weak
boson scattering. Therefore a corroborating analysis should be possible in the WBF production mode, although it will require a careful
event selection in order to isolate the Higgs-related contribution from electroweak production of $Z$-pairs and jets.

While we have focused on $H \to ZZ$ exclusively, we expect qualitatively similar effects in $H \to W^+W^-$ decay.
However, this decay channel is experimentally challenging in the presence of a jet, due to the large top-pair background.
We also note that Higgs results in the exclusive one-jet bin are known to be sensitive to the logarithms of the 
transverse momentum veto \cite{Stewart:2011cf}. Moult and Stewart recently showed \cite{Moult:2014pja} that although
these logarithms have a mild impact on the Higgs width measurement in $H \to ZZ$ decay, where the study is rather inclusive
in the number of jets, they can have a large effect in the $H\to W^+W^-$ channel where jet-binning is crucial to the analysis.
The LO study that we present in this work should be taken as a starting point, and higher order corrections should be computed
when at all possible.

\section*{Acknowledgements}
RKE would like to thank the RWTH, Aachen for hospitality during
the preparation of this paper and would like to acknowledge useful discussions with Michael Czakon
and Sebastian Kirchner. 
RR is grateful to the CERN Theory Group for their hospitality during the preparation of this paper.
This research is supported by the US DOE under contract DE-AC02-07CH11359.

\appendix
\section{Definition of the scalar integrals}
\label{Intdef}
We work in the Bjorken-Drell metric so that
$l^2=l_0^2-l_1^2-l_2^2-l_3^2$. 
The definition of the integrals is as follows
\begin{eqnarray}
&& B_0(p_1;m_1,m_2)  =
 \frac{\mu^{4-D}}{i \pi^{\frac{D}{2}}\rG}\int d^D l \;
 \frac{1}
{(l^2-m_1^2+i\varepsilon)
((l+p_1)^2-m_2^2+i\varepsilon)}\,,\nn \\
&& C_0(p_1,p_2;m_1,m_2,m_3)  =
\frac{1}{i \pi^2}
\nn \\
&& \times \int d^4 l \;
 \frac{1}
{(l^2-m_1^2+i\varepsilon)
((l+p_1)^2-m_2^2+i\varepsilon)
((l+p_1+p_2)^2-m_3^2+i\varepsilon)}\,,\nn \\
&&\nn \\
&&
D_0(p_1,p_2,p_3;m_1,m_2,m_3,m_4)
=
\frac{1}{i \pi^2}
\nn \\
&&
\times \int d^4 l \;
 \frac{1}
{(l^2-m_1^2+i\varepsilon)
((l+p_1)^2-m_2^2+i\varepsilon)
((l+p_1+p_2)^2-m_3^2+i\varepsilon)
((l+p_1+p_2+p_3)^2-m_4^2+i\varepsilon)}\,, \nn \\
\end{eqnarray}

We have removed the overall constant which occurs in $D$-dimensional integrals
\beq
\rG\equiv\frac{\Gamma^2(1-\e)\Gamma(1+\e)}{\Gamma(1-2\e)} =
\frac{1}{\Gamma(1-\e)} +{\cal O}(\e^3) =
1-\e \gamma+\e^2\Big[\frac{\gamma^2}{2}-\frac{\pi^2}{12}\Big]
+{\cal O}(\e^3)\,.
\eeq
\bibliography{H+jet}
\end{document}